\documentclass[draft]{agujournal2018}
\usepackage{apacite}
\usepackage{url} 
\usepackage{lineno}
\nolinenumbers
\addeditor{EC}
\addeditor{AS}

\journalname{Geochemistry, Geophysics, Geosystems}

\begin{document}

%
%


\title{Possible orthopyroxene enrichment in the upper mantle below the
Mississippi Embayment}

%
%




\authors{Arushi Saxena\affil{1}, Eunseo Choi
\affil{1}, Christine A. Powell\affil{1}, 
Charles A. Langston\affil{1}}


\affiliation{1}{Center for Earthquake Research and Information, The University of Memphis}




\correspondingauthor{Arushi Saxena}{asaxena@memphis.edu}




\begin{keypoints}
  \item Possible origins of similar-magnitude and negative ($-$ 5 to 6\%) Vp and Vs anomalies at depths 100 to 250 km beneath the NMSZ are explored
  \item A combination of elevated temperature and water content cannot explain Vp and Vs anomalies
  \item The Vp and Vs anomalies can be explained by an orthopyroxene content of upto 40\% along with temperature variations
\end{keypoints}

%
%


\begin{abstract}
A high-resolution tomography study for the mantle beneath the New Madrid
Seismic Zone (NMSZ), a major intraplate earthquake zone in the Central
and Eastern US, reveals 3 - 5 \% low Vp and Vs anomalies in the upper
mantle in the depth range 100 to 250 km. When attributed only to
temperature variations, such low velocities lead to temperatures higher
than the olivine solidus, for which consistent observational support is
lacking. Similar magnitudes of the Vp and Vs anomalies also suggest that
temperature anomalies are unlikely to be the sole factor because~Vs is
more sensitive to temperature than Vp. In this study, we attribute the
velocity anomalies to elevated water and orthopyroxene (Opx) contents as
well as temperature variations. We then compute differential stresses
using three-dimensional numerical models subjected to a loading similar
to the regional stresses. The models assume a Maxwell viscoelastic crust
and mantle with viscosities based on the temperature, water and Opx
content converted from the tomography. We find that the presence of
water allows for sub-solidus variations in temperature. 
However, any combination of water content and
temperature anomalies fails to yield Vp and Vs anomalies of a similar
magnitude. Thus, we consider Opx enrichment in place of water content.
Our calculations show that reasonable Opx content and sub-solidus
temperatures can explain both Vp and Vs anomalies. 
We speculate that the Opx enrichment could have been produced by fluids released from a stagnant piece of the Farallon slab imaged at around 670 km beneath the NMSZ.
\end{abstract}

%
%

%


%
%
%
%
\section{Introduction}
{\label{424512}}

\begin{figure}[ht]
\centering
\includegraphics[width=0.70\textwidth]{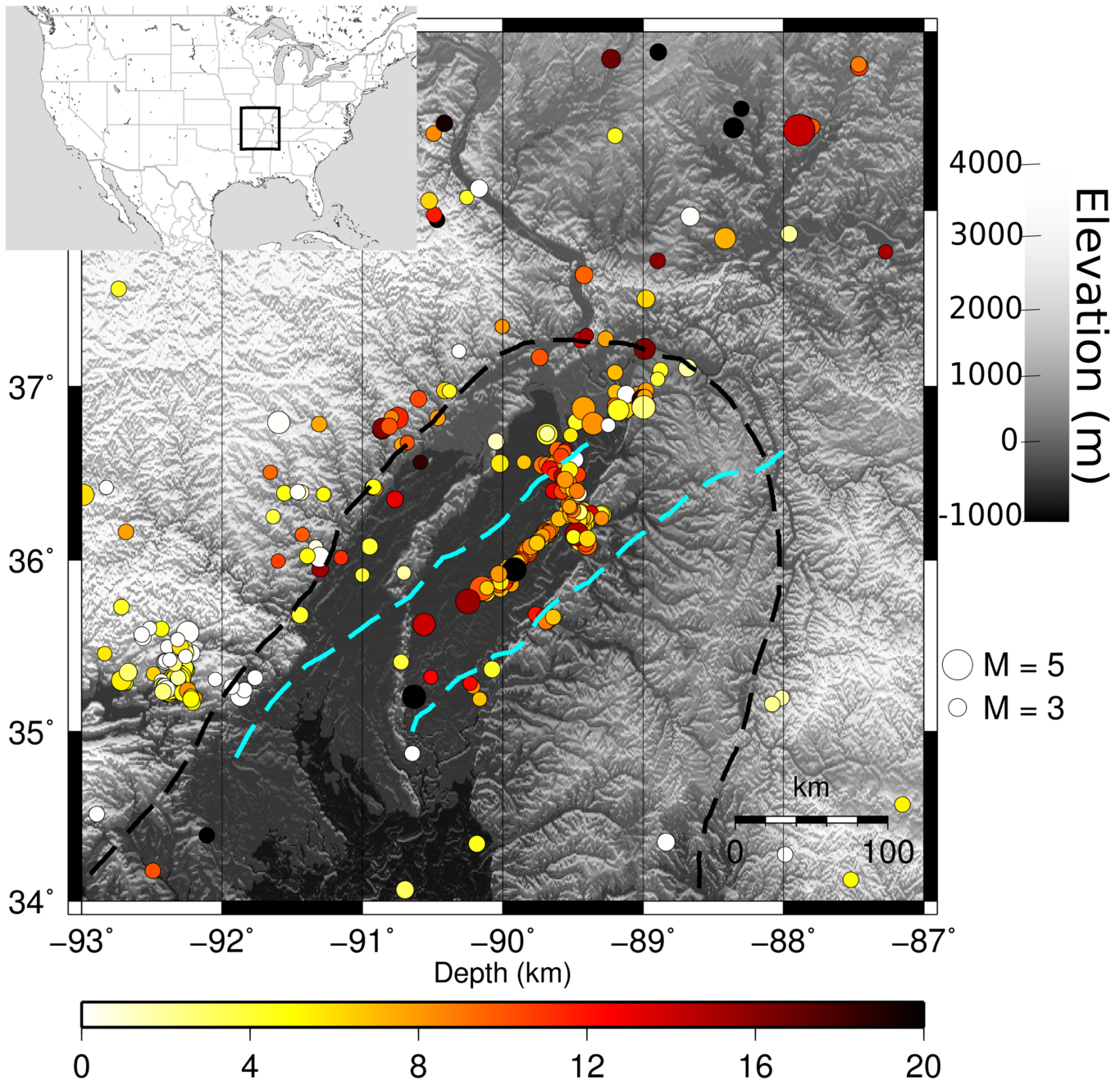}
\caption{{Earthquake epicenters (\textgreater{}Mw2.0) in the New Madrid Seismic
Zone (NMSZ) from May, 2002 to May, 2017. The color and size of symbols
represent the depth and magnitude, respectively. The black line marks
the boundary of the Mississippi Embayment and the Reelfoot Rift is
contained within the blue dashed line. The inset shows a part of the
North American and the U.S. state borders providing a geographic
reference for the location of the NMSZ (black square). The earthquake
catalogue is obtained from the United States Geological Survey
at \url{https://earthquake.usgs.gov/earthquakes/search/}. This figure is
available under~CC-BY~\citet{powell2018b}.
\label{621864}
}}
\end{figure}

The New Madrid Seismic Zone (NMSZ), located in the Northern Mississippi
Embayment, is a seismically active region as indicated by paleoseismic
evidence~\citep[e.g.,][]{Tuttle_2002} and three historic large (Mw
$>$ 7) earthquakes in 1811- 1812~\citep{Johnston_1996,Bakun_2004}. Numerous
small-magnitude earthquakes (Mw $<$ 5) are still being recorded
(Fig. {\ref{621864}}) posing a potential risk of
seismic hazard in and around this zone.

Geodetic strain rates observed in this region are less than 3 $\times$ 10$^{-9}$/yr~\citep{Boyd_2015}. This value is much
lower than the strain rate~loading of active plate boundaries suggesting
that the earthquake generation is not entirely associated with plate
tectonics. This has~puzzled researchers, invoking them to come up with
alternative mechanisms to explain the seismicity of NMSZ that are
compatible with the field observations.

The existing models for the NMSZ seismicity cannot explain all of the
existing seismic and other geophysical observations, calling for an
improved model. Previous studies that proposed lower lithospheric
strength~\citep{Liu_1997}, weaker lower
crust~\citep{Kenner_2000a}, deglaciation of the Laurentide ice
sheet~\citep{Grollimund_2001}, sinking of a denser heterogeneous lower
crust~\citep{Pollitz_2001}, and higher mantle
temperatures~\citep{Zhan_2016} in the NMSZ than in the surrounding
region would elevate the stress level and therefore promote seismicity.
However, although required to be weak by some models~\citep{Liu_1997,Kenner_2000a}, the lower crust in the
NMSZ is strong due to the presence of a dense mafic intrusion (commonly
called a pillow)~\citep{Mooney_1983,Rabak_2011} possibly created by the passage of
the Bermuda hotspot during mid-Cretaceous~\citep{Cox_1997,Cox_2002,Chu_2013}.
Attributing the seismic speed variations in the region's upper mantle
only to anomalous temperature~\citep{Zhan_2016}~also seems
inconsistent with more recent high resolution~tomographic
observations~\citep{Chen_2016,Chen_2014,Nyamwandha_2016}.~\citet{Chen_2014}~and~\citet{Chen_2016}
found 5 \% Vp and 7 \% Vs decrease at depths of 75-150 km beneath the
NMSZ, respectively. When converted to temperature variations only,
seismic velocity anomalies of this magnitude would mean partial melting
in the upper mantle. ~This is not supported by surface heatflow
observations. Since Vs is more sensitive to temperature~\citep[e.g.,][]{Goes_2000,Cammarano_2003} and the presence of melt alters Vs more than
Vp~\citep{Karato_2003}, Vs and Vp anomalies of a similar magnitude
(3\% -5\%) observed by~\citet{Nyamwandha_2016} cannot be explained by
temperature variations alone.

We note that two recent regional tomographic studies in the Central and
Eastern United States (CEUS)~\citep{Nyamwandha_2016,Chen_2016} independently
attributed~their observed low velocity anomalies not only to elevated
temperature but also to other factors such as the presence of water and
compositional variations. The presence of orthopyroxene (Opx) lowers Vp
more than Vs~\citep{Schutt_2010} and provides a possible explanation
for the similar magnitudes of the Vp and Vs anomalies in the presence of
elevated temperature. Previous studies have reported Opx enrichment over
flat subducted slabs~\citep[e.g.,][]{Wagner_2008,Tang_2012}.

We propose that Opx enrichment occurs from the flat Laramide slab below
the NMSZ (\citealt{Sigloch_2008}) at mantle transition depths , altering
the composition and hydrating the above lying lithosphere. ~Similar
reasoning is invoked to explain the tomographic and geochemical
observations in the North China Craton (NCC)~\citep[e.g.,][]{Zhang_2009,Santosh_2010,Wang_2014} in
which asthenospheric upwellings from the stagnant pacific slab altered
the lithosphiere.

To address this idea in our study, we constrain the upper mantle
viscosity beneath the NMSZ assuming that the low velocity anomalies
found in the~\citet{Nyamwandha_2016} tomography results come from
temperature, water and Opx. We compute differential stresses at
seismogenic depths using the viscosity distributions as an input for
geodynamic numerical models and include the effects of compositional
variations accounting for possible enrichment in Opx.

\section{Modeling Methods}
\subsection{Viscosity Calculations}

\subsubsection{Case I: Temperature only~}
{\label{433817}}

Effective viscosity for power law creep is given as~\citep{Kirby_1987}: 
\begin{equation} \label{eq1}
 \eta_{eff} = \dot{\varepsilon} ^{\frac{1-n}{n}} A^{-1/n} \exp \left(-\frac{H}{nRT} \right),
\end{equation}
where $\dot{\varepsilon}$ is the effective strain rate taken as $3.25 \times 10^{-17} s^{-1}$ for our calculations \cite{Calais_2006}, $R$ is the gas constant, $A$ is the pre-exponential factor, $n$ is the power law exponent, $H$ is the activation energy and $T$ is the temperature. The values of $A$, $n$ and $H$ as well as density for each layer are listed in Table~\ref{table1}.  . 
%
\begin{table}[ht]
\caption{{Rheological constants used in our model simulations }} \label{table1}

\centering
 \begin{tabular}{ c  c  c  c  c  c } 
 \hline
 Layer  & Composition & Density (kg/m$^3$) & A (MPa$^{-n}$s$^{-1}$) & n & H (kJ mol$^{-1}$) \\ 
 \hline
 Upper crust$^a$  & Quartzite & 2750 & 1.1 $e^{-4}$  & 4.0 & 223  \\ 
 Lower crust$^a$  & Granulite & 2900 & 1.4 $e^4$  & 4.2 & 445 \\
 Mantle$^b$  & Olivine & 3200 & 1.1 $e^5$ & 3.5 & 530 \\ 
 \hline
\end{tabular} 
  \begin{tablenotes}
     $a:$ \citet{Burov2010}, $b:$ \citet{Dixon_2004}
  \end{tablenotes}
\end{table}

\begin{figure}[ht]
\centering
\includegraphics[width=0.7\textwidth]{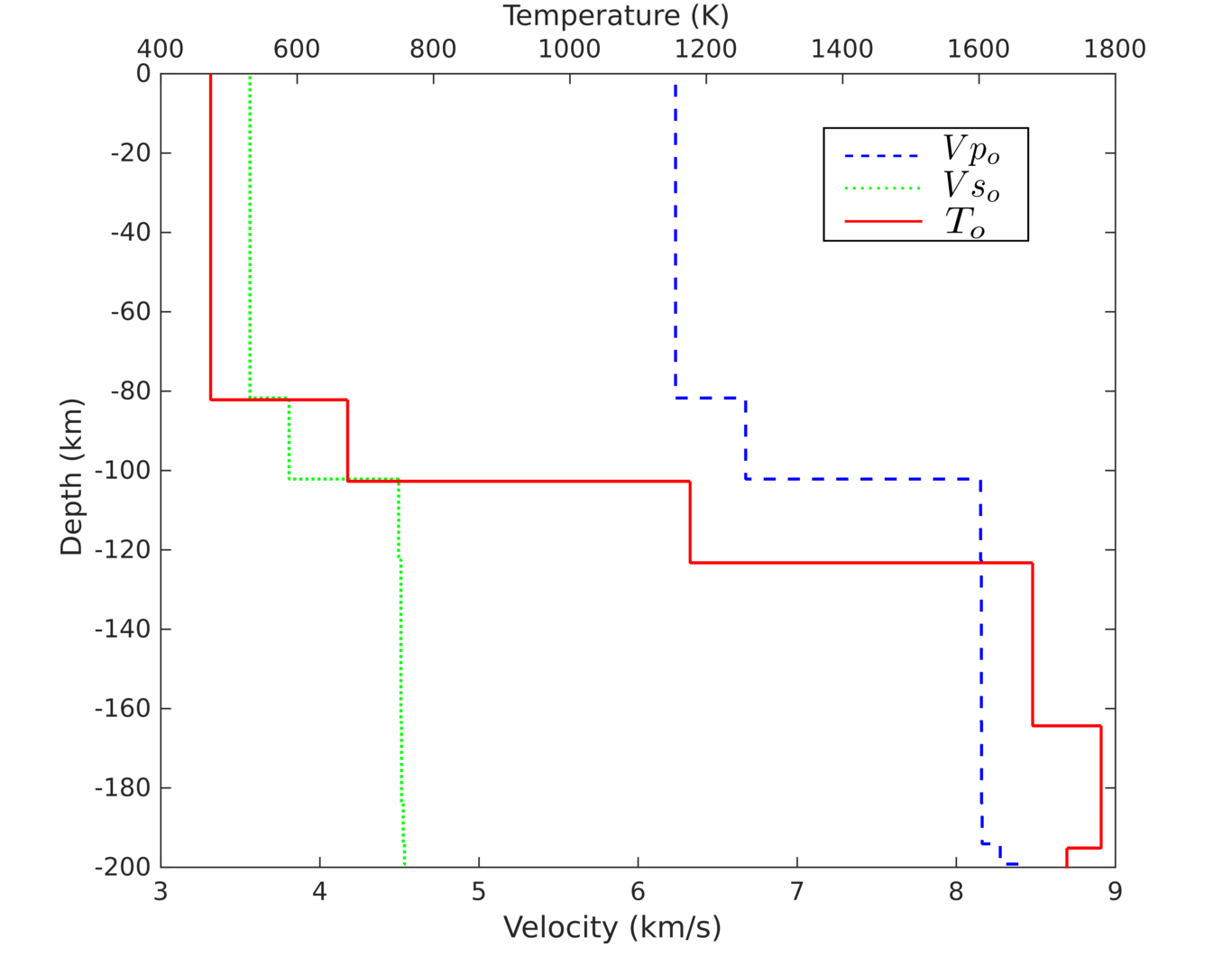}
\caption{{Reference P and S wave
velocities ($Vp_{0}$ and $Vs_{0}$) used by~\citet{Nyamwandha_2016} for their tomography results and reference temperature ($T_{0}$) from~\citet{Liu_1997} and~\citet{Goes_2002} based on which the temperature anomalies are
calculated. This figure is available under~CC-BY~\protect\citet{powell2018b}.
{\label{236955}}
}}
\end{figure}

We calculate temperature anomalies using the Vp and Vs anomalies,
denoted as~\(\partial Vp\) and~\(\partial Vs\) hereafter,
determined by~\citet{Nyamwandha_2016}. 
The sensitivities of Vp and Vs to temperature are based on the anelastic and
anharmonic effects in olivine (Ol) and are originally given
in~\citet{Goes_2000} as follows: 
\begin{eqnarray*}
 \frac{\partial Vp}{\partial T} & = &  \frac{1}{2 Vp_{0} \ \rho} \frac{\partial K}{\partial T} + \frac{2}{3 Vp_{0} \ \rho} \frac{\partial \mu}{\partial T} +  Q_p^{-1} \frac{aH}{2 R T_{0}^2 \tan(\pi a/2)}, \label{eq2} \\
 Q_p & = &  A \omega^{a} \exp \left[ \frac{a(H+PV)}{RT_{0}} \right] \frac{3Vp_{0}^{2}}{4Vs_{0}^{2}}, \nonumber \\
 \frac{\partial Vs}{\partial T} & = &  \frac{1}{2 Vs_{0}\,  \rho} \frac{\partial \mu}{\partial T} + Q_s^{-1} \frac{aH}{2\, R\, T_{0}^{2} \tan(\pi a/2)}, \label{eq3} \\ 
 Q_s & = & A \omega^{a} \exp \left[ \frac{a(H+PV)}{RT_{0}} \right]. \nonumber
\end{eqnarray*}
Density is assumed to be constant as its
sensitivity to temperature is about 4 orders of magnitude smaller than
those of the elastic moduli when the standard value of volume expansion
coefficient (10$^{-5}$/K) is assumed.
Table~{\ref{table2}} lists the values of parameters and
relevant quantities. The reference velocities
(Vp$_{0}$, Vs$_{0}$) and the reference temperature
(T$_{0}$) are depth-dependent
(Fig.~{\ref{236955}}). We use the~velocity model
from~\citet{Pujol_1997} for the crustal depths ($<$ 40 km) and
IASP91~\citep{Kennett_1991} for the mantle ($>$ 40 km). The
reference temperatures are taken from Table 2 in~\citet{Zhan_2016},
which are originally from~\citet{Liu_1997} for depths shallower than
100 km and from~\citet{Goes_2002} for greater depths.
\begin{table}[ht]
\caption{{Elastic moduli, their temperature derivatives and anelasticity parameters  }}
\label{table2}
\centering
 \begin{tabular}{ l  c  c  c } 
 \hline
  Variable & Symbol & Unit & Value \\
  \hline
  Bulk modulus$^{a}$ & $K$ & GPa & 129 \\ 
  Shear modulus$^{a}$ & $\mu$ & GPa & 82 \\ 
  Derivative with temperature$^{a}$ & $\partial K / \partial T$ & MPa/K & $-$16 \\ 
   & $\partial \mu / \partial T$  & MPa/K & $-$14 \\ 
  Angular frequency$^{b}$ & $\omega$ & rad/s & 2$\pi$ \\ 
  Frequency exponent$^{a,c}$ & $a$ & N/A &  0.15 \\ 
  Gas constant & $R$ & J/mol$\cdot$K & 8.314 \\ 
  Activation energy$^{a,c}$ & $H$ & kJ/mol & 500 \\ 
  Volume$^{a,c}$ & $V$ & $cm^3$/mol & 20 \\ 
  \hline
\end{tabular}
  \begin{tablenotes}
    $a$: \citet{Goes_2000}, $b$: based on the frequency range of the tomography by \citet{Nyamwandha_2016}, $c$: \citet{Sobolev_1996}.
  \end{tablenotes}
\end{table}

\subsubsection{Case II: Temperature and
water} \label{396366}

Velocity anomalies are assumed to be a function of water content as well as temperature in this case. For simplicity, we assume uniform distributions of water that are completely independent of velocity and temperature variations. The effective viscosity is derived from the power law creep under the constant strain rate assumption~\citep{Dixon_2004}:
\begin{equation}
\eta_{eff} = \dot{\varepsilon}^{\frac{1-n}{n}} C_{OH}^{-\frac{r}{n}} A^{-\frac{1}{n}} \exp \left(-\frac{H}{nRT} \right).
\label{dixon}
\end{equation}
where $r$ is the fugacity exponent, assumed to be 1.2~\citep{Hirth_2003}, $C_{OH}$ is water content in Ol given in the unit of H/10$^6$ Si. Under the current assumption, the temperature $T$ for (\ref{dixon}) is the reference temperature, $T_{0}$ from Fig.~\ref{236955}. Parameters for the upper and lower crust are listed in Table~\ref{table1} and those for the mantle are taken from~\citet{Hirth_2003}.

We consider two possible scenarios for water concentration in olivine
following~\citet{Dixon_2004}: Damp (50 ppm~ H$_{2}$O) and wet
(fully saturated with water). Although it is not realistic to assume
that the water distributions are uniform throughout our model domain, we
explore these scenarios to assess the viscosity-lowering effects of the
water content.

\subsubsection{Case III: Temperature and water added
selectively} {\label{249721}}

\begin{figure}[ht]
\centering
\includegraphics[width=0.5\textwidth]{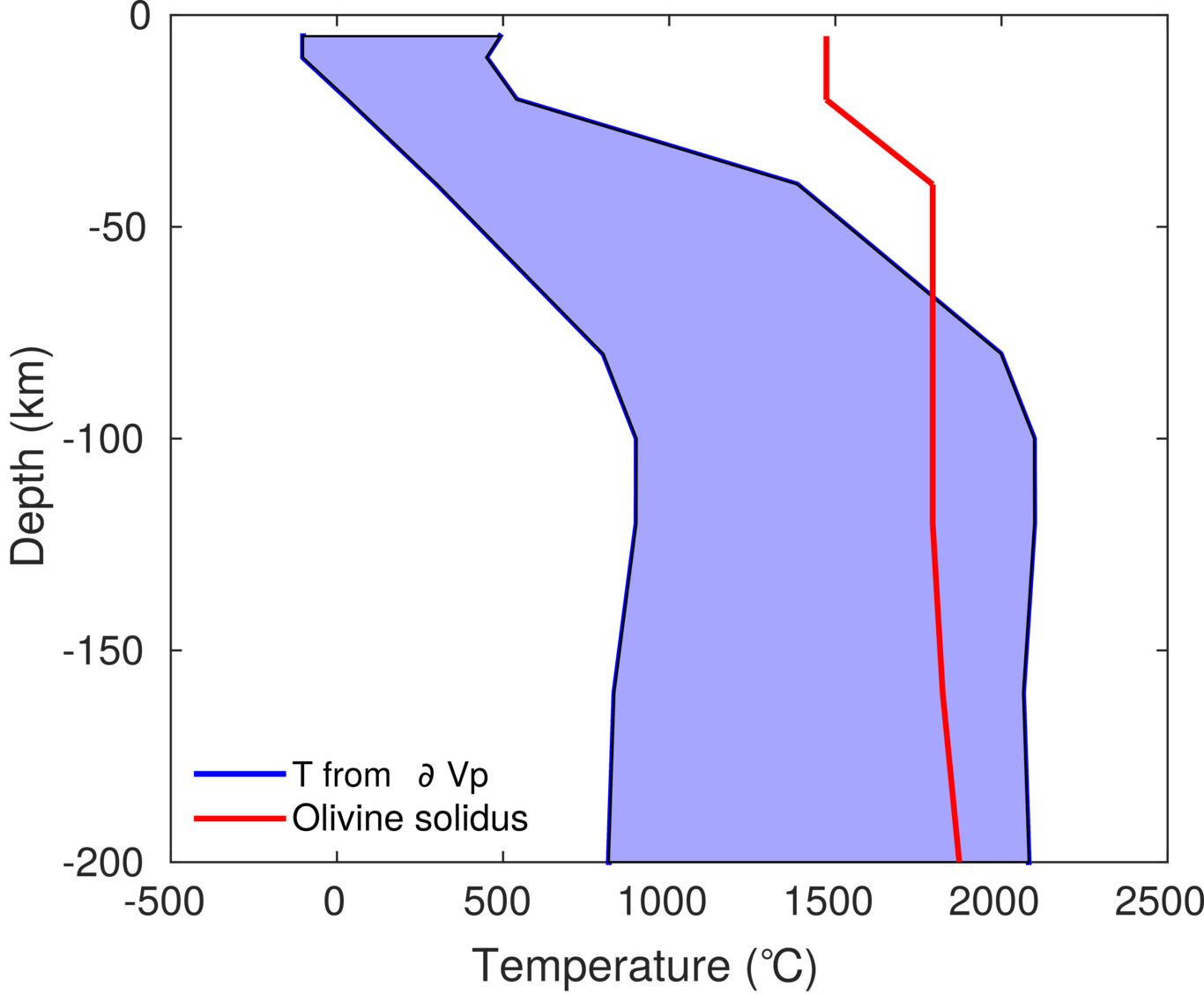}
\caption{{Minimum and maximum temperatures (blue lines) calculated from P wave
anomalies~as in Case I; and the solidus (red line)
from~\citet{Herzberg_1983}. This figure is available
under~CC-BY~\protect\citet{powell2018b}.
{\label{982076}}%
}}
\end{figure}

 We assume that mantle temperatures are likely to be lower than the olivine solidus ($T_{m}$) everywhere in the study area. This is because the presence of melt would alter Vs more than Vp~\citep{Karato_2003} but the region's tomography results show similar magnitudes of $\partial Vp$ and $\partial Vs$. If computed as in Case I, temperatures ($T_{I}$) would exceed $T_{m}$ at some locations (Fig. \ref{982076}). Only at those locations, we attribute the excess temperature  ($T_{ex} = T_{I}-T_{m}$) to the presence of water such that $\partial Vp = \partial Vp^{T_m} + \partial Vp^{OH}$, where $\partial Vp^{T_m}$ is the Vp anomaly due to $T_{m}-T_{0}$ and $\partial V{p}^{OH}$ is contribution to the Vp anomaly from the water. We use the following relationship between fluid content and velocity anomaly~\citep{Karato_1998}:
\begin{equation} 
     \delta C_{OH} = - \frac{ \partial V^{OH}}{V_o (T,P)} \left( 2 \tan \frac {\pi \alpha}{2} \right) \frac {\omega}{\alpha B} \left(\frac{A}{\omega}\right)^{1-\alpha} \exp \left(\alpha \beta \frac{T_m}{T_{0}}\right), \label{eq5}
\end{equation}
where  $\partial V^{OH}$ is $\partial Vp - \partial Vp^{T_m} = \partial Vp - \partial V{p}/ \partial T (T_{m} - T_{0})$ , $T_m$ is the Ol solidus and $T_{0}$ is the reference temperature. $V_{0}(T,P)$ is the reference velocity, $\omega = 2\pi$ rad/s (Table \ref{table2}),  $\beta$ is related to enthalpy $H$ (in Table~\ref{table1}) as $H = \beta R T_m$~\citep{Karato_1998}, $A = B = 8.8 \times 10^{6}$ and $\alpha = 0.3$, which corresponds to the half space cooling model from \citet{Karato_1998}. $\delta C_{OH}$ is calculated in ppm H/Si assuming that the reference state has no water content i.e. $C_{OH} = 0$.

\subsubsection{Case IV: Temperature and Opx} {\label{646083}}

Opx enrichment, together with temperature variations, may explain the
$\partial Vp$ and $\partial Vs$ of similar magnitudes because
Vp is known to be more sensitive to Opx concentration than Vs~\citep{Schutt_2010} while Vs is more sensitive to temperature than Vp.
We therefore consider the combined effects of Opx and temperature on Vp
and Vs anomalies as follows:
\begin{equation}
\partial V_{k} = \frac{\partial V_{k}}{\partial T} \, \delta T (X_{Opx})+ \frac{\partial V_{k}}{\partial X_{Opx}} \, \delta X_{Opx}, \label{eq6}
\end{equation}
where $V_{k}$ denotes Vs or Vp, $X_{Opx}$ is the Opx volume fraction in vol \% and $T$ is temperature (K). The coefficients are defined by the Voigt scheme: i.e. in case of Vs,
\[  \frac{\partial V_s}{\partial T} = \frac{\partial V_s}{\partial \mu}\frac{\partial \mu}{\partial T} =  \frac{1}{2V_s\,\rho} \frac{\partial}{\partial T} \left[ (1-X_{Opx})\,\mu_{Ol} + X_{Opx}\,\mu_{Opx} \right], \] where $\rho$ is the reference density, $\mu$, $\mu_{Ol}$ and $\mu_{Opx}$ are shear moduli for the bulk composition, pure Ol and pure Opx, respectively. The sensitivities of elastic moduli to temperature for Ol and Opx are taken from~\citet{Goes_2000} and the sensitivities of Vp and Vs to Opx content from~\citet{Schutt_2010}. We invert the above equations for $X_{Opx}$ and $\delta T$ simultaneously using the Newton-Raphson scheme assuming that Opx concentration is zero in the reference state.

There is no wide consensus on whether the presence of Opx would
generally strengthen or weaken the mantle~\citep[e.g.,][]{McDonnell_2000,Ji_2001,Sundberg_2008,Tikoff_2010,Tasaka_2013,Hansen_2015}). We calculate bulk viscosity in the presence of Opx
using a constant strain scheme as $\eta=(1-X_{Opx})\eta_{ol}+X_{Opx}\eta_{opx}$~\citep[e.g.,][]{Ji_2001}.
In one model, we assume that Opx viscosity
is 3.3 times that of Ol~\citep{Hansen_2015} for strengthening
effects and in another we construct a model in which Opx viscosity 
is one eighth of Ol viscosity~\citep{Ji_2001} to account for 
weakening effects of Opx.

\subsection{Differential Stress Calculations} {\label{589841}}

\begin{figure}[ht]
\begin{center}
\includegraphics[width=1.0\textwidth]{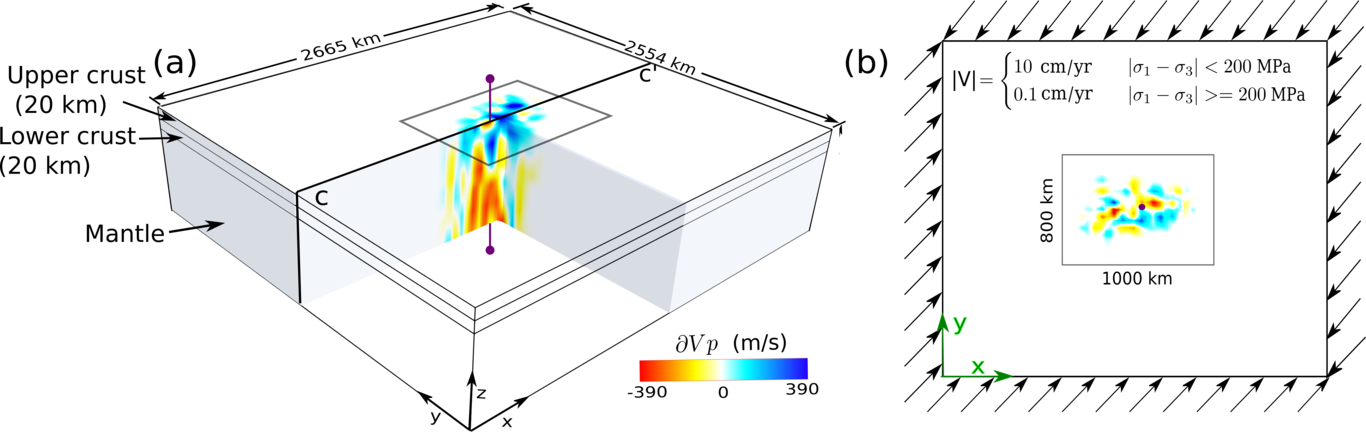}
\caption{{(a) Model domain and the P wave anomalies from~\protect\citet{Nyamwandha_2016}. The
domain has 3 layers: 20 km-thick upper crust, 20 km-thick lower crust
and mantle. 
The purple vertical line through
the NMSZ marks the location of vertical profiles of model results shown
in the subsequent figures. The rectangle on the surface represent the
spatial extent of the tomography model. (b) Cross-section of the model
domain and S wave anomalies from~\protect\citet{Nyamwandha_2016} at 100 km depth.
Arrows denote the velocity boundary conditions applied on the side
walls.~This figure is available under~CC-BY~\citet{powell2018b}.
{\label{869833}}%
}}
\end{center}
\end{figure}

We construct three-dimensional (3D) models with Maxwell linear viscoelastic rheology to compute differential stresses. Each model has three layers: upper crust (0-20 km), lower crust (20-40 km,) and mantle (40-200 km) (Fig.~\ref{869833}a). The lateral extent of our model is 2665 $\times$ 2554 km. The central region (gray box) is constrained by the ~\citet{Nyamwandha_2016} tomography results(Fig.~\ref{869833}a) but the overall domain is set to be much greater than that of the tomography in order to reduce boundary effects.

To simulate northeast-trending regional compressional principal stress ($\sigma_1$) ~\citep{Zoback_1989}, horizontal velocities with a magnitude of 10 cm/yr are applied on the X- and Y-axis-perpendicular faces (Fig.~\ref{869833}b) until the differential stress ($|\sigma_{1}-\sigma_{3}|$) of 200 MPa is achieved at the depth of 10 km~\citep{Zoback_1993,Baird_2010}. After that, the boundary velocity magnitude is reduced to avoid further increasing the differential stress but is set to a non-zero value, 0.1 cm/yr, to maintain this stress state without complete viscous relaxation. The top surface is free (i.e., zero-traction) and the bottom has a no-slip condition.

We discretize the domain into a mesh with 227448 hexahedral tri-linear elements using Trelis 15.0. The mesh has lateral and vertical resolutions of 25 and 5 km, respectively. We use Pylith~\citep{Aagaard2013}, an open source finite element code for crustal dynamics developed and distributed by Computational Infrastructure for Geodynamics, to calculate stresses and displacements under the loading conditions described above. Differential stresses are derived from stress solutions at one location in the NMSZ marked by the purple line in Fig.~\ref{869833}a.

\subsection{Caveats regarding model setup} {\label{722743}}

We assume that the reference state for the tomography has uniformly zero
Opx concentration. When a non-zero initial Opx concentration is
assumed, the Opx change calculated from inversion is reduced by the
initial amount assumed, keeping the total Opx concentration the same. The inverted temperatures are also reduced by 20 K if we
assume an initial Opx concentration of 20 \%, although we do not show the
results with different initial opx concentration.

Also, although the magnitude of the temperature anomaly is reduced when
water is present, the temperature due to~our maximum calculated water
content in Case III ($\sim$40 ppm H/Si) changes by
$\sim$1 \% from the original value~\citep{Karato_2003}, which
we neglect in our calculations. We explore Case II only to observe the
weakening effects on viscosity due to the presence of water, therefore,
it is acceptable to consider temperature independently from water
content.

Our model setup does not include the effects of gravity and it is
assumed that the density of all 3 layers, upper crust, lower crust, and
mantle, is constant within each layer. Since we are interested only in
the effects of local mantle heterogeneity on the overlying crust,
gravity in our model setup can be neglected. Additionally, ignoring
density contrasts due to temperature anomalies is justifiable as the
volumetric expansion coefficient for common rocks is $\sim$
$10^{-5}$/K and for maximum temperatures calculated here, 600 K, this
amounts to 0.6\% change in the density. We also ignore the effects of
density on composition~in Eq. 6 and take the bulk density as 3220
kg/m3~using the average of the densities of Ol (3222 kg/m$^{3}$) and Opx
(3218 kg/m$^{3}$). This assumption is reasonable as the densities of Ol and
Opx (taken from~\citet{Goes_2000}) are similar and using average
density introduces a maximum change of 0.06 \% compared to the case where
density is calculated by the weighted average.

We also assess the effects of the Reelfoot Rift (Fig.~\ref{621864}) by
imposing $-$6 \% Vp and Vs anomalies in the rift~\citep{Pollitz_2014} modeled as $\sim$70 km wide on the
surface, narrowing to $\sim$35 km at the depth of 10 km~\citep{Marshak_1996}. With the imposed low velocity anomalies,
viscosities within the rift are lower than the surrounding areas and
therefore, the resultant differential stresses from the model with the
rift are lower than the models without rifts. This observation is
consistent with the effects of rift zones found by~\citet{Zhan_2016}. We do not show the model results with the rift as we are
interested in explaining the cause of the low-velocity anomaly
($>$50 km depths) and it's impact on the NMSZ seismicity. The
presence of rifts lowers the absolute differential stress in the upper
crust by $\sim$10 MPa but does not change the relative
impact on differential stress among the cases investigated in this
study.

\section{Model Results} {\label{334878}}
\subsection{Case 1: Temperature only} 
\begin{figure}[ht]
\centering
\includegraphics[width=1.0\textwidth]{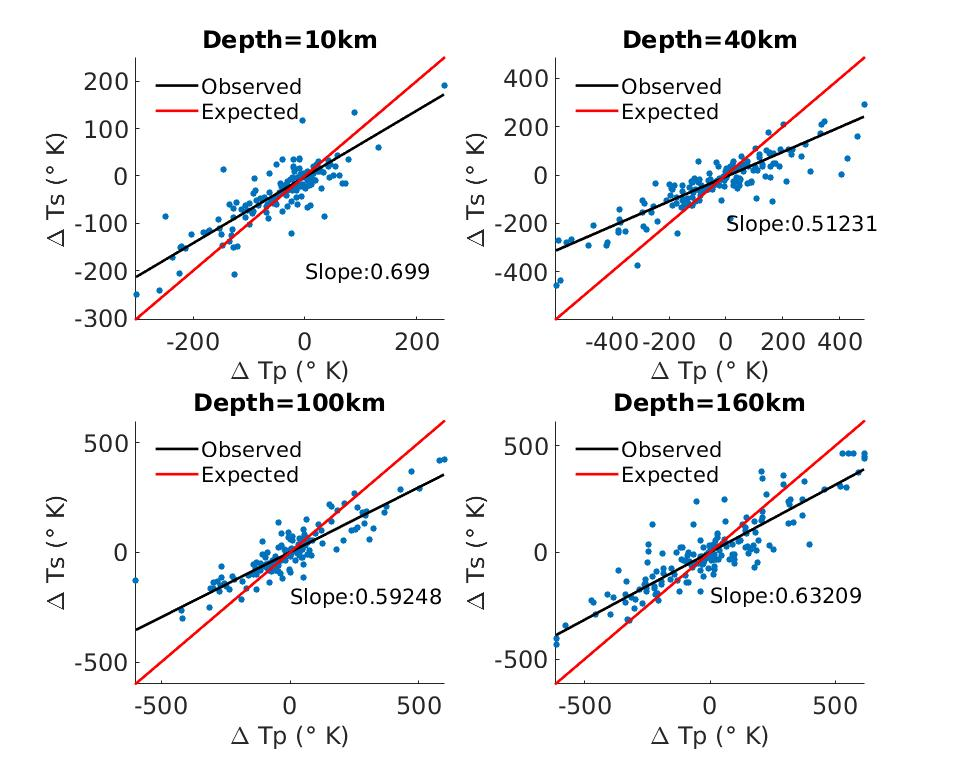}
\caption{{Scatter plots of temperature anomalies calculated from P wave ($\Delta T_{p}$) and S wave ($\Delta T_{s}$) anomalies with the
best-fitting lines (black). The red line has a slope equal to 1,
representing the case where both Vp and Vs anomalies are due to
temperature only. This figure is available
under~CC-BY~\citet{powell2018b}.
{\label{943687}}%
}}
\end{figure}

{\label{661403}}
We convert $\partial Vp$ and $\partial Vs$ to temperature variations using (\ref{eq2}) and (\ref{eq3}). Temperature anomalies are plotted  in Fig. \ref{943687} at different depths with best-fitting linear trends (black solid lines, Fig. \ref{943687}). If both $\partial Vp$ and $\partial Vs$ are simultaneously explained by temperature only, the converted temperatures should be equal falling on a line with the slope equal to 1 (red lines, Fig. \ref{943687}) and the magnitudes of $\partial Vs$ would be greater than those of $\partial Vp$ because Vs has greater sensitivity to temperature than Vp. However, 
slopes of the best-fitting lines are in the range 0.5 to 0.7 at all depths, consistently smaller than one because $\partial Vp$ and $\partial Vs$ have similar magnitudes in the tomography by~\citet{Nyamwandha_2016}.

\begin{figure}[ht]
\centering
\includegraphics[width=0.70\textwidth]{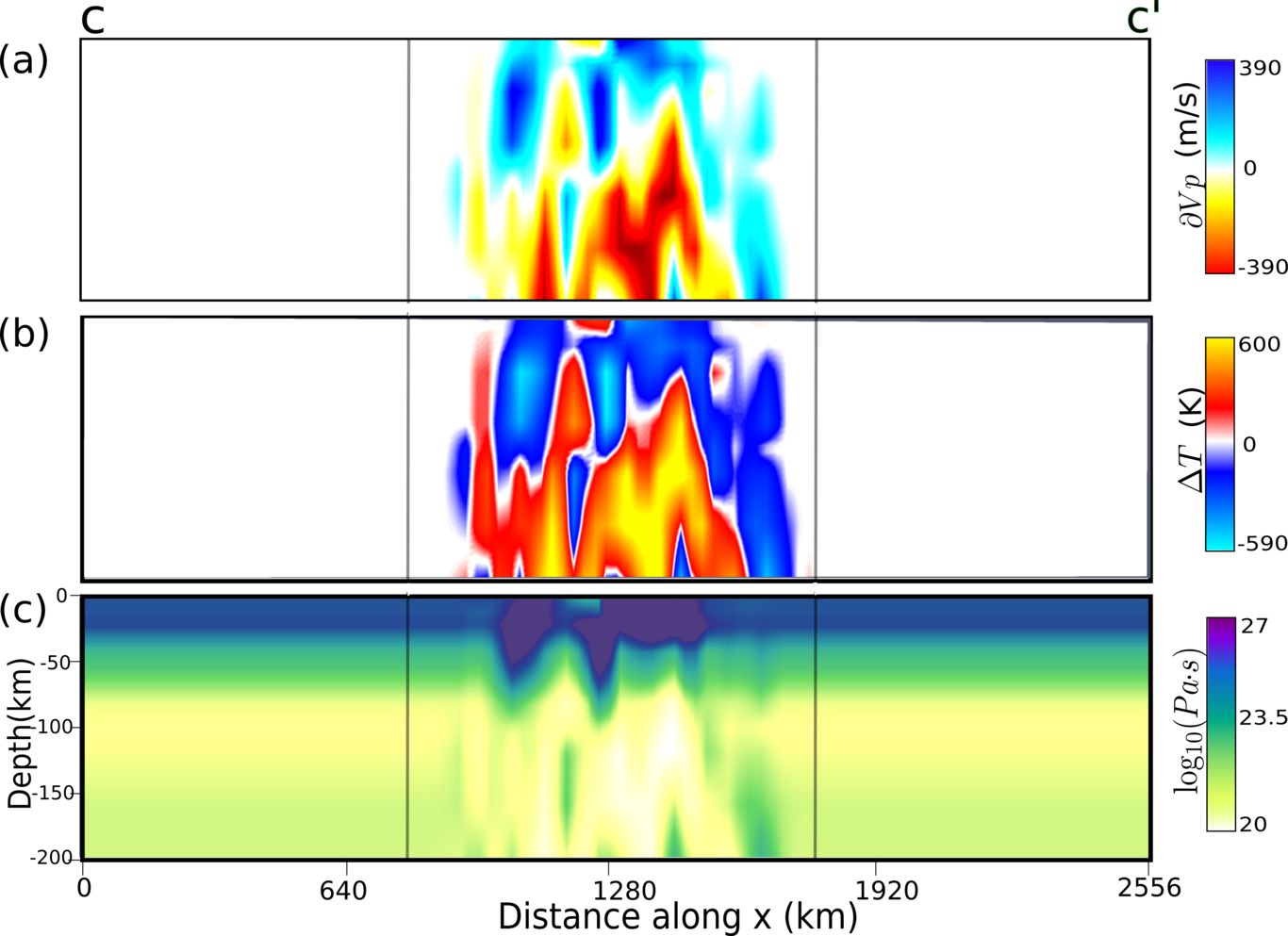}
\caption{{Conversion of the velocity anomalies into model viscosities at
cross-section CC' in Fig.~{\ref{869833}}a. Thin
vertical gray lines mark the boundaries of the tomography domain
of~\citet{Nyamwandha_2016} (see
Fig.~{\ref{869833}}a).~Vertical axis is exaggerated by
a factor of 3. (a) P-wave velocity anomalies ($\partial V_{p}$) from
~\citet{Nyamwandha_2016}. (b) Temperature anomalies calculated based
on $\partial V_{p}$. (c) Viscosities computed based on the
temperatures anomalies in (b) and the reference temperature shown in
Fig. {\ref{236955}}. This figure is available
under~CC-BY~\citet{powell2018b}.
{\label{834796}}%
}}
\end{figure}

We show one of the viscosity calculations based on the
$\partial Vp$-converted temperature anomalies at cross-section
C-C$^{\prime}$ (see Fig.~\ref{869833}) in Fig.~\ref{834796}.
The Vp anomalies from \citet{Nyamwandha_2016} along this
cross-section are shown in Fig.~\ref{834796}a. According to
the Vp sensitivity to temperature, (\ref{eq2}), and
parameters in Table~\ref{table2}, the temperature variations
are in the range $\pm$600 K (Fig.~\ref{834796}b). We compute
the viscosity using (\ref{eq1}) and the total temperature,
which is the sum of the reference geotherm (Fig.~\ref{236955})
and the temperature variations. The viscosity distribution
(Fig.~\ref{834796}c) follows the pattern of velocity and
temperature variations and the minimum is about 10$^{20}$
Pa$\cdot$s where Vp is the smallest.

\subsection{Case II: Temperature and uniform water
content} \label{case-ii-temperature-and-uniform-water-content}

\begin{figure}[ht]
\centering
\includegraphics[width=1.0\textwidth]{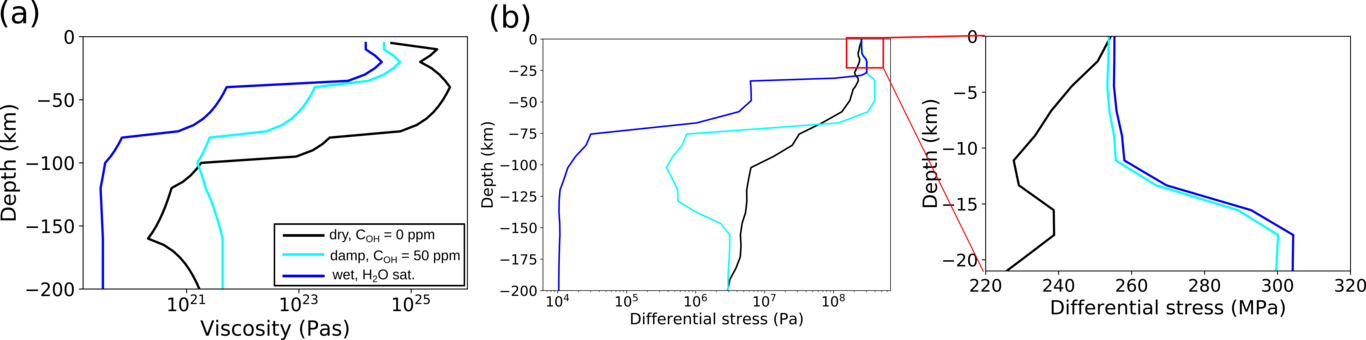}
\caption{{Depth profiles of (a) viscosity~(b) differential stresses~ within the
NMSZ (purple line in Fig.~{\ref{869833}}) for the model
computed from Vp-based temperatures calculated in Case I (black),~
reference temperatures from Fig.~{\ref{236955}}~with
uniform C\textsubscript{OH} concentration of 50 ppm (cyan), reference
temperatures from Fig.~{\ref{236955}}~with
water-saturated olivine concentration (blue). Differential stresses (b)
are zoomed into to observe the concentration in upper crust on the
right.~This figure is available under~CC-BY~\citet{powell2018b}.
{\label{322886}}%
}}
\end{figure}

We compute differential stresses within the NMSZ based on the reference temperature (Fig.~\ref{236955}) and a spatially-uniform concentration of $H_2 O$ in olivine: 50 ppm (damp) and water saturated concentration (wet) (Fig.~\ref{322886}a). We compare our calculations with the viscosity computed from $\partial Vp$-converted temperatures as in Case I and no water content, referred as the dry case ($C_{OH} = 0$ ppm) subsequently. Given the same temperature, water can reduce differential stress in the upper mantle (depths $\ge$ 40 km) by orders of magnitude but in the upper crust, differential stress increases with water content by about 2 MPa (Fig.~\ref{322886}b). The same effect of water content on differential stress level in the upper crust is seen in the models based on Vs anomalies (not shown here).

\subsection{Case III: Temperature and selectively-added water}
{\label{965647}}

\begin{figure}[ht]
\centering
\includegraphics[width=1.0\textwidth]{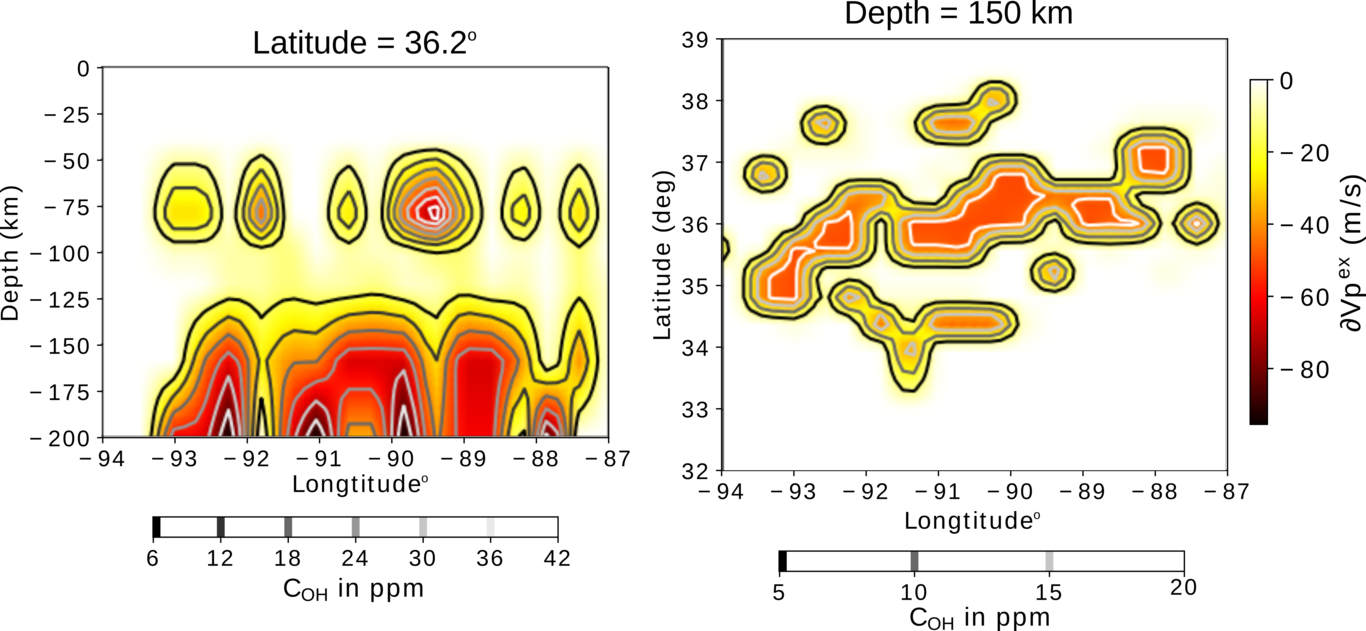}
\caption{{Selectively-added water contents based on the velocity anomalies
converted from the temperatures in the super-solidus region shown on a
cross-section along the latitude of 36.2\(^{\circ}\) (left), which
approximately coincides with CC' in Fig.~{\ref{869833}}
and on a depth section at 150 km (right). The lower velocities
correspond to higher calculated water contents.~This figure is available
under~CC-BY~\citet{powell2018b}.
{\label{157690}}%
}}
\end{figure}

We compute water content in the super-solidus regions
identified in Case I (Fig. \ref{982076}) using (\ref{eq5})
and plot the contours of the calculated water contents on the
Vp anomalies calculated from the temperatures in the
super-solidus region, $\partial V_{p}^{ex}$, at a depth 150 km
and in the cross section (marked in Fig.~\ref{869833}) in
Fig.~\ref{157690}. The correlation between the calculated
water content and low $\partial V_{p}$ is clearly seen in the
cross section and the depth section (Fig.~\ref{157690}). For
our calculations from the velocity anomalies (\ref{eq5}),
water contents reach up to 50 ppm in low-Vp regions.

\begin{figure}[ht]
\centering
\includegraphics[width=1.0\textwidth]{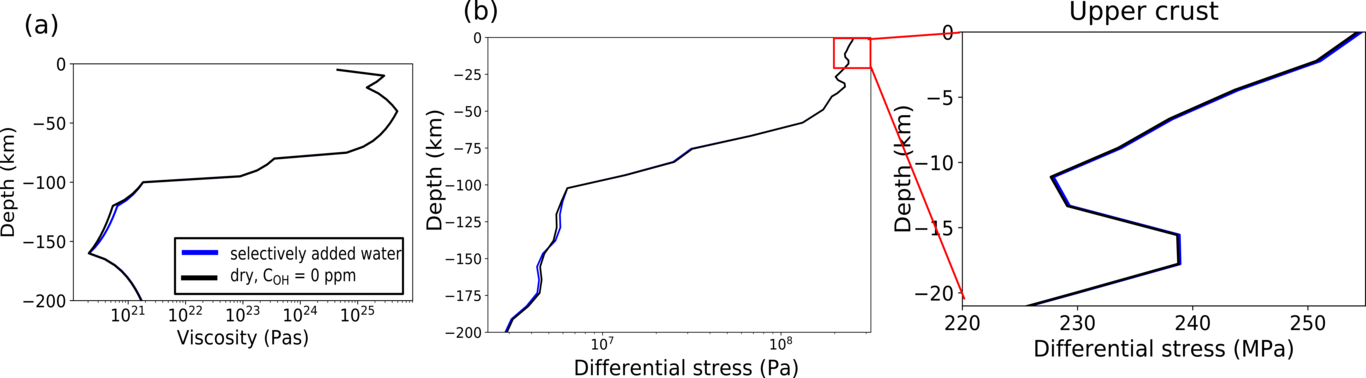}
\caption{{Same as Fig.~{\ref{322886}}~but for the model with
selectively-added~ water in blue and the dry model~
described in Case I in black.~This figure is available
under~CC-BY~\citet{powell2018b}.
{\label{273212}}%
}}
\end{figure}

Based on the water content only in the supersolidus regions, we calculate the viscosity distribution using (\ref{dixon}) and the parameters in Table~\ref{table1} within the NMSZ (Fig. \ref{869833}). The resultant differential stresses in the NMSZ upper crust are compared with the dry model (Case I) in Fig.~\ref{273212}b. A minor decrease in viscosity due to the  weakening effect of the water is seen only in the depth range from 120 to 160 km, where significant water content is calculated (Fig.~\ref{273212}a) . Differential stresses also show only minor differences between the dry model and the present model over the same depth range (Fig.~\ref{273212}b).

\subsection{Case IV: Temperature and Opx content}
{\label{923350}}

\begin{figure}[ht]
\centering
\includegraphics[width=1.0\textwidth]{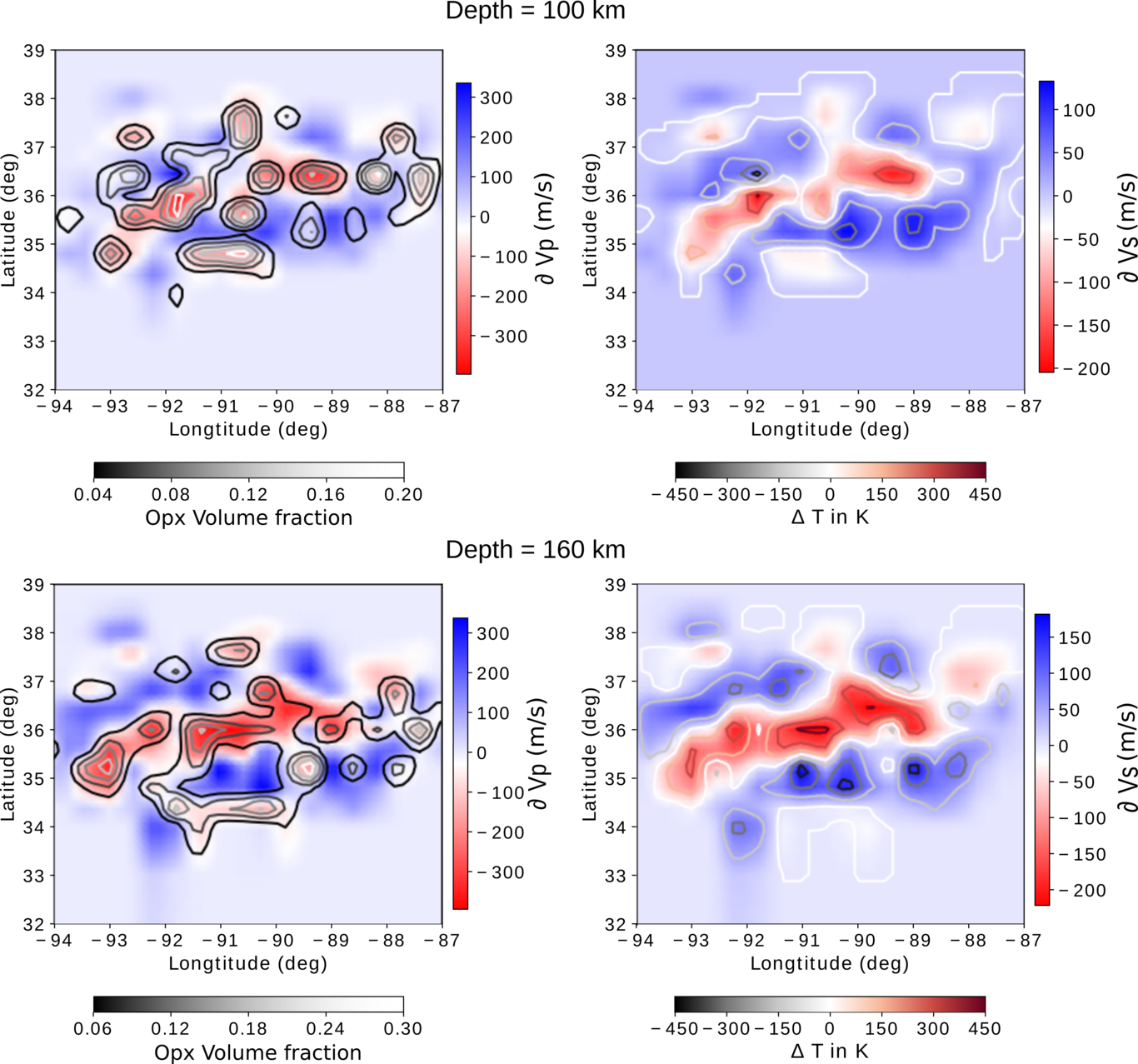}
\caption{{Inverted Opx contents and temperature anomalies at depths of 100 and 160
km. (left) Vp anomalies overlain by the contours of the Opx volume
fraction.~(right)~~Vs anomalies overlain by the contours of the
temperature variations. This figure is available
under~CC-BY~\citet{powell2018b}.
{\label{994587}}%
}}
\end{figure}

We compute Opx volume fraction ($X_{Opx}$) and temperature anomalies ($\Delta T$) that simultaneously satisfy $\partial Vs$ and $\partial Vp$ by solving the system of equations (\ref{eq6}). Two depth slices at 100 and 160 km displaying $X_{Opx}$ and $\Delta T$ are plotted in Fig.\ref{994587}. Since $X_{Opx}$ is more sensitive to Vp and $\Delta T$ is more sensitive to Vs, we plot $X_{Opx}$ and Vp together in the left panels and $\Delta T$ and Vs together in the right panels in Fig.\ref{994587}. $X_{Opx}$ values show correlation with negative Vp anomalies, reaching a maximum of 0.2 at 100 km depth and 0.3 at 160 km. Temperature anomalies vary between $\pm$450 K. This magnitude of temperature anomalies is 10-20 \% smaller than that of the dry, Opx-free case (Case I), which exhibited temperature anomalies of $\pm$600 and $\pm$500 K at the corresponding depths (Fig.\ref{834796}).

\begin{figure}[ht]
\centering
\includegraphics[width=1.0\textwidth]{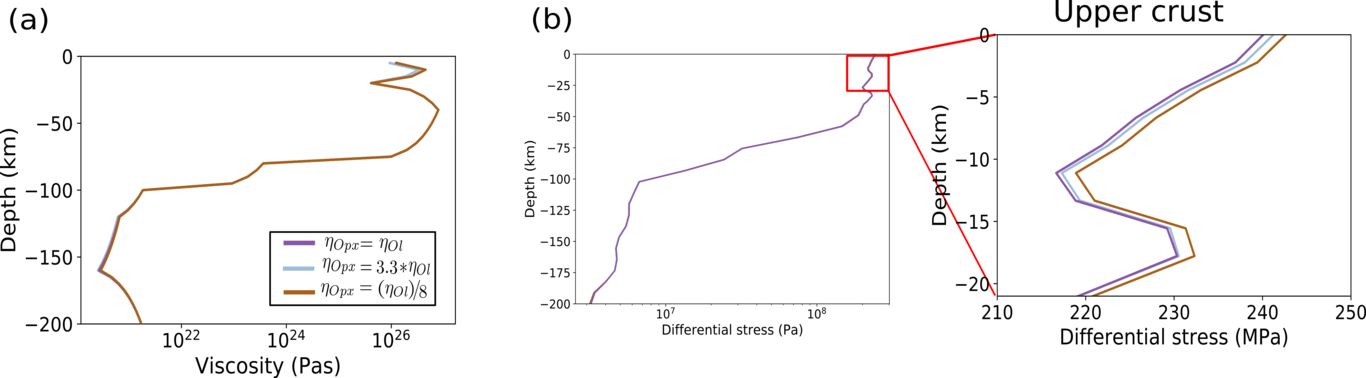}
\caption{{Same as Fig.~{\ref{322886}}~but for the models
corresponding to the three possible effects of Opx enrichment on
viscosity: Viscosity strengthening (light blue), viscosity weakening
(brown) and the same viscosity as olivine (purple). This figure is
available under~CC-BY~\citet{powell2018b}.
{\label{322217}}%
}}
\end{figure}

Differential stresses are computed under various assumptions on the effects of Opx enrichment on viscosity (Fig.~\ref{322217}). We consider three possibilities: Opx enrichment increases viscosity \cite{Hansen_2015}, decreases it \cite{Ji_2001}, or has no effect on viscosity. Estimates for the sensitivity of viscosity to Opx contents are only loosely constrained~\citep[e.g.,][]{McDonnell_2000,Ji_2001,Tikoff_2010,Tasaka_2013,Hansen_2015}. \citet{Hansen_2015} suggest a factor of 1.2 to 3.3 increase in viscosity for a pyroxene fraction upto 30 \%. Conversely, \citet{Ji_2001} indicate a factor of 3 to 8 decrease in the strength of enstatite, the Mg end member of Opx, relative to fosterite, the Mg end member of Ol, for samples containing 0.4-0.6 volume fraction of fosterite. We choose a factor of 3.3 for increase and 8 for the decrease in Opx viscosity relative to Ol as end member cases. Our calculations for these three cases show that the strengthening and weakening effects by Opx enrichment are insignificant (Fig.~\ref{322217}a). Differential stresses also show only about 1 \% of variation at all depths (Fig.~\ref{322217}b). 
%

\section{Discussion} {\label{781975}}

\begin{figure}[ht]
\centering
\includegraphics[width=1.0\textwidth]{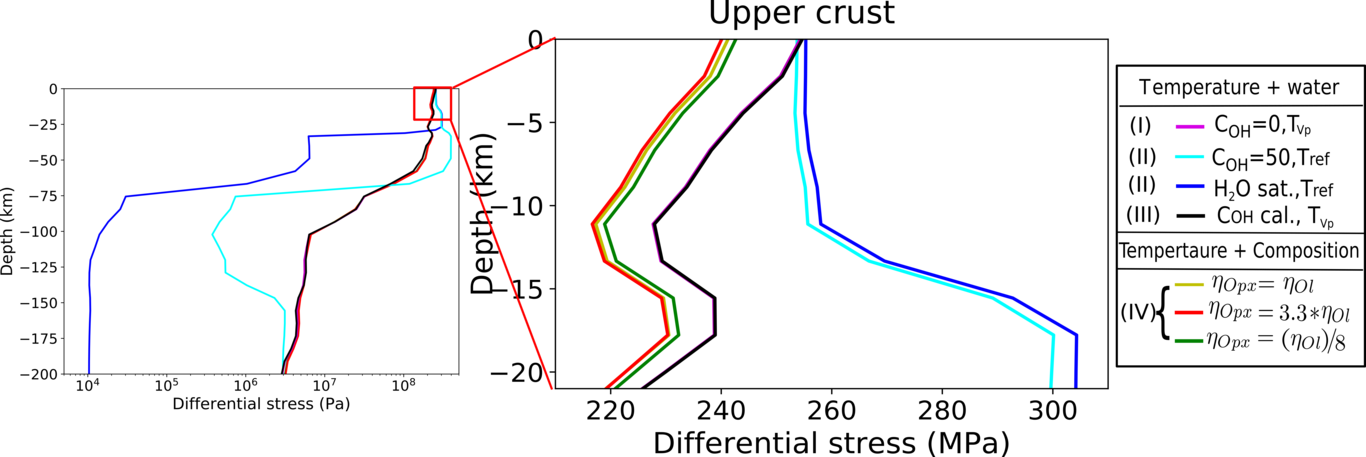}
\caption{{Depth profiles of differential stress from all the cases (Case I to IV)
presented in this study. Case I: temperatures converted from
the $\partial V_{p}$ (purple); Case II: water added in the whole
domain with the reference temperature (cyan and blue); Case III:~
selective addition of water in super-solidus region and temperature
calculated from Case I using $\partial V_{p}$ (black); Case IV:
temperatures inverted using $\partial V_{p}$
and $\partial V_{s}$ and Opx for its strengthening (red), weakening
(green) and no (yellow) viscosity effects on Opx. This figure is
available under~CC-BY~\citet{powell2018b}.
{\label{538935}}%
}}
\end{figure}

Differential stress calculated in all the cases described 
in the study is most sensitive to the temperature and less so to
the Opx content and water content. Low differential stresses 
in the mantle in all the models (cases I-IV) are consistent with
the high-temperature anomalies and low
mantle viscosity values. Relatively strong crust (depths $>$
40 km) concentrates the differential stress from the deeper mantle.
Differential stress in the upper crust calculated from the
Opx-enriched cases (case IV) appears to be~ smaller than values
computed using forward calculated temperatures (case I and II).
This trend is the
result of lower temperatures (Fig.~{\ref{994587}})
observed after inversion in case IV than the calculated temperatures in
case I (Fig.~{\ref{943687}}), implying that the mantle
in case I is weaker and concentrates higher differential stress in the
upper crust (Fig.~{\ref{538935}}). It can also be
observed from Fig.~{\ref{538935}} that the effects of
Opx on differential stress are minor. Capping the maximum temperatures
computed from P wave anomalies ($\Delta T_{p}$) with the olivine
solidus and attributing the excess anomalies to~ water content~ (case
III) has minor effect on the differential stress compared to case I. On
the other hand, the differential stress calculated using uniform water
content (case II) and the reference temperature (Tref) differs
appreciably from all the other cases (I, III) using
the $\partial V_{p}$-calculated temperatures. Therefore it can be
inferred from Fig.{\ref{538935}} that the differential
stress is most sensitive to the temperature.

Nonetheless, we favor the differential stress calculations with the
inverted temperatures and the presence of Opx (Case IV) because~only
they can explain the similar magnitudes of the upper mantle Vp and Vs
anomalies determined by~\citet{Nyamwandha_2016}. 
The velocity anomalies cannot be attributed to elevated
temperature alone, a more fertile mantle, or to the presence of fluids
combined with elevated temperatures; the presence of elevated amounts of
Opx is a necessary condition. The inverted values of Opx in the range of
5-45 \% (Fig. {\ref{994587}}) are similar to those found
previously for the Kaapvaal Craton~\citep{Wagner_2008} and the Colorado
Plateau~\citep{Li_2008}. Metasomatism by Si rich fluids derived
from a subducting slab is the preferred mechanism for Opx enrichment in
these regions as documented by the texture and composition of mantle
xenoliths~\citep{Smith_1999,Bell_2005,Li_2008}.

The need for Opx enrichment to provide a reasonable explanation of the
similar Vp and Vs anomalies argues for the presence of a slab below the
Mississippi Embayment at some point in time. Flat subduction of the
Farallon slab is usually invoked to explain deformation and magmatism
during the Laramide orogeny (ca. 80 to 35 Ma) in the western U.S.
\citep[e.g.,][]{Humphreys_2009} and several studies suggest that flat slab
segments are present at depths above 1000 km in the central U.S.~\citep{Sigloch_2008,Liu_2008,Sigloch_2011,Schmandt_2014,Gao_2014,Porritt_2014}.
Flattening of the Farallon slab can be attributed to the subduction of
oceanic~ plateaus (the Shatsky and Hess plateaus) located on an old
($>$50 My) slab~\citep[e.g.,][]{English_2004,Liu_2010}. This situation would
result in cold slab subduction, facilitating transport of hydrous
minerals to transition zone
depths~\citep{Maruyama_2007,Kusky_2014,Wang_2018}.
Geodynamic modeling by ~\citet{Liu_2010} indicates that the top of
the Hess plateau is located at a depth of about 660 km below the
Mississippi Embayment (Figure 4 in ~\citep{Liu_2010}).

Seismic velocity models determined by tomographic inversion of
teleseismic body waves also indicate the presence of high velocity
regions at or below the transition zone in the central U.S. that are
interpreted as fragments of the Farallon slab~\citep[e.g.,][]{Sigloch_2008,Sigloch_2011,Schmandt_2014,Porritt_2014}.
All of these models indicate the presence of a high velocity region in
the vicinity of the Mississippi Embayment with, in general, shallower
high velocities to the east of the NMSZ. The P and S wave velocity
models determined by~\citet{Porritt_2014} indicate the presence of high
velocities in the depth range 200 to 800 km below the Embayment and
surrounding area. In the depth range 400 to 600 km, the high velocity
region is split by a band of low velocities that trends NNE, roughly
along the western boundary of the Embayment. The low velocity band is
either not present or cannot be resolved below 600 km. The low velocity
band corresponds to the low velocity region imaged in the depth range
100 to 250 km below the Embayment in~ the higher resolution study
by~using TA and NELE FlexArray stations.

\citet{Biryol_2016} attribute the high velocity upper mantle region
located to the east of the NMSZ to the presence of a dense, partially
removed, lithospheric drip. The down-going drip may result in upward,
return flow of the asthenosphere below the Reelfoot rift (Figure 9
in~\citep{Biryol_2016}). Our analysis indicates that the Vp and Vs
anomalies determined
by~\citet{Nyamwandha_2016} cannot be attributed to the presence of asthenosphere
devoid of Opx enrichment.~ One possibility that may be compatible with
the concept put forth by~\citet{Biryol_2016} is that the upwelling
asthenosphere was enriched in Opx by the presence of a slab fragment
that has subsequently dropped to depths just below the transition zone.
A deeper slab fragment is suggested by the tomography results of~\citet{Schmandt_2014} and by the present location of the Hess oceanic
plateau determined by~\citet{Liu_2010}.

\section{Conclusions} {\label{195931}}

In this study, we explore possible explanations for the~ upper mantle Vp
and Vs anomalies ($\partial V_{p}$ and $\partial V_{s}$), negative
and of similar magnitudes, found below the Mississippi Embayment in the
tomographic study by~\citet{Nyamwandha_2016} and suggest Opx enrichment as
the most plausible one. When inverted for Opx volume fraction and
temperature anomalies, these low velocity anomalies~correspond to
positive temperature anomalies up to 450 K~and Opx contents of up to
40 \%. The presence of Opx reduces the range of temperature variations,
which is $\pm$ 600 K when converted from $\partial V_{p}$
only. The estimated Opx fraction is in the range of estimates made for
the Kaapvaal Craton~\citep{Wagner_2008} and the Colorado Plateau~\citep{Li_2008}, which are associated with the water-rock
interactions from the Chile--Argentina and Farallon flat slab,
respectively. In NMSZ, olivine alteration to Opx might have been induced
by fluids released from the flat-subducted Laramide slab.

We set up numerical viscoelastic models based on the tomography
by~\citet{Nyamwandha_2016} to assess how variations in temperature, water and
Opx contents affect the regional stress field in three dimensions. The
concentration of stress from the weaker mantle into the stronger
overlying crust has been indicated previously to explain the seismicity
of the NMSZ~\citep[e.g.,][]{Kenner_2000a,Zhan_2016}. We observe
similar stress-concentrating effects due to warm upper mantle indicated
by the low velocity anomalies.

\acknowledgments
 We thank C. Nyamwandha for access to the tomography data. 






%
%
%
%
%
%
%
%
%
%

\bibliography{bibliography/biblio} 


\end{document}